# Evaluating Effectiveness of Tamper-proofing on Dynamic Graph Software Watermarks

Prof. Dr. Malik Sikandar Hayat Khiyal
Department of Computer Science and Software Engineering
Fatima Jinnah Women University
Rawalpindi, Pakistan
.

Aihab Khan
Department of Computer Science
Fatima Jinnah Women University
Rawalpindi, Pakistan
.

Dr. M. Shahid Khalil
Department of Mechanical Engineering
University of Engineering and Technology
Texila, Pakistan
.

Sehrish Amjad
Graduate, Department of Computer Science
Fatima Jinnah Women University
Rawalpindi, Pakistan
.

*Abstract*—**For enhancing the protection level of dynamic graph software watermarks and for the purpose of conducting the analysis which evaluates the effect of integrating two software protection techniques such as software watermarking and tamper-proofing, constant encoding technique along with the enhancement through the idea of constant splitting is proposed. In this paper Thomborson technique has been implemented with the scheme of breaking constants which enables to encode all constants without having any consideration about their values with respect to the value of watermark tree. Experimental analysis which have been conducted and provided in this paper concludes that the constant encoding process significantly increases the code size, heap space usage, and execution time, while making the tamper-proofed code resilient to variety of semantic preserving program transformation attacks.** *(Abstract)*

## I. Introduction

The most significant property involved in digital information is that, it is in principle very easy to produce and distribute unlimited number of its copies. This might undermine the music, film, book and software industries and therefore it brings a variety of important problems. In this paper the dilemma under consideration is of software piracy, concerning the protection of the intellectual and production rights that badly need to be solved.

To overwhelm the problem of software piracy through watermarking, the emerging technique is dynamic graph software watermarking. But the major drawback involve in this technique is the lack of stealthiness. To protect dynamic graph software watermarks against attacks due to lack of resemblance between the watermark code and source program, constant encoding technique of tamper-proofing dynamic graph software watermarks was first proposed by Yong He in [17] and then with few modification again proposed by Clark Thomborosn in [16].This novel tamper-proofing method is based on encoding constant of software programs into data structure of watermark tree by the means of protecting against program transformation attacks.

Section I of the paper comprised on basic introduction of the problem found in different already proposed constant encoding techniques. In Section II, tamper-proofing techniques of dynamic graph software watermarks along with their limitations are discussed. While Section III provides the understanding of the structure. Section IV deals with the implemented technique. In Section V, outcome of this paper i.e. experimental results are mentioned and several suggestions for further enhancement are described in Section VI.

### A. Contribution

This paper provides the details about the practical experience with the idea of constant encoding techniques, and then also measures the effect of incorporating two software protection techniques which are watermarking and tamper-proofing. The purpose of the underlying effort is to find out the proportion of change due to combining two software protection techniques on different parameters such as code size, execution time, heap space usage and resilience.

## II. Related Work

To tamper-proof the dynamic and static software watermarks, different methods are employed. But in this paper, our main focus is on the tamper-proofing of Dynamic Graph Watermarks. In our literature search, we find only a few publications which describe the tamper-proofing method of Dynamic Graph Watermarks. The only method which is described through out the publications is of constant encoding. Constant encoding method evaluates from the idea of Palsberg described in [21]. In constant encoding, the main concept which is followed is to replace the constant values with function calls.

In [17], Yong He described the constant encoding technique for tamper-proofing the DGW. The basic idea behind this technique is the replacement of constant values with function calls. The value return by the function calls is dependent upon the values of pointer variables in the dynamic data structure of same shape as the watermark tree. The graph structure utilized by this technique is of planted plane cubic tree (PPCT) shape. The procedure utilizes by the Yong technique for encoding the constants successfully helps in creating the false dependencies from the watermarked program to the constant tree structure. The distinguishing property of Yong's technique is handling of large constant due to not having the constraint on the size of the constant tree. But unfortunately the effectiveness of this protection technique depends upon the assumption that the attacker is unable to differentiate between the watermark building code and constant tree building code.

In [16], Clark Thomborson proposed the constant encoding technique as the modification of Yong He technique. To overcome the weakness of existence of false dependency between the watermark tree and the constant tree, the idea of Thomborson is based on utilization of watermark tree for finding substructure instead of separately creating the constant tree. The concept provided by the Thomborson reinforce the Yong's constant encoding technique in this way that even if the attacker successfully locates the watermark, he will not able to remove or modify it without taking risk of changing the constant values.

### III. FRAMEWORK OVERVIEW

#### A. Framework/Model of Constant Encoding Technique in Collaboration of Constant Splitting Technique

Procedure of constant encoding is performed on the watermarked program generated by the SandMark system by selecting the subsequent options.

- *Watermark Type: Numeric*
- *Storage Method: Hash*
- *Storage Policy: Formal*
- *Protection Method: if: safe: try*
- *GraphType:SandMark.util.newgraph.codec.PlantedPlaneCubicTree*
- *Use Cycle Graph: No*
- *Sub graph Count: 1*
- *Inline Code: No*
- *Replace Watermark Class: No*
- *Dump Intermediate Code: No*

After watermarking, decompilation is performed to convert generated file of byte code into java source code. Constant encoding technique implemented in this paper handles constant of only one type which is of numeric.

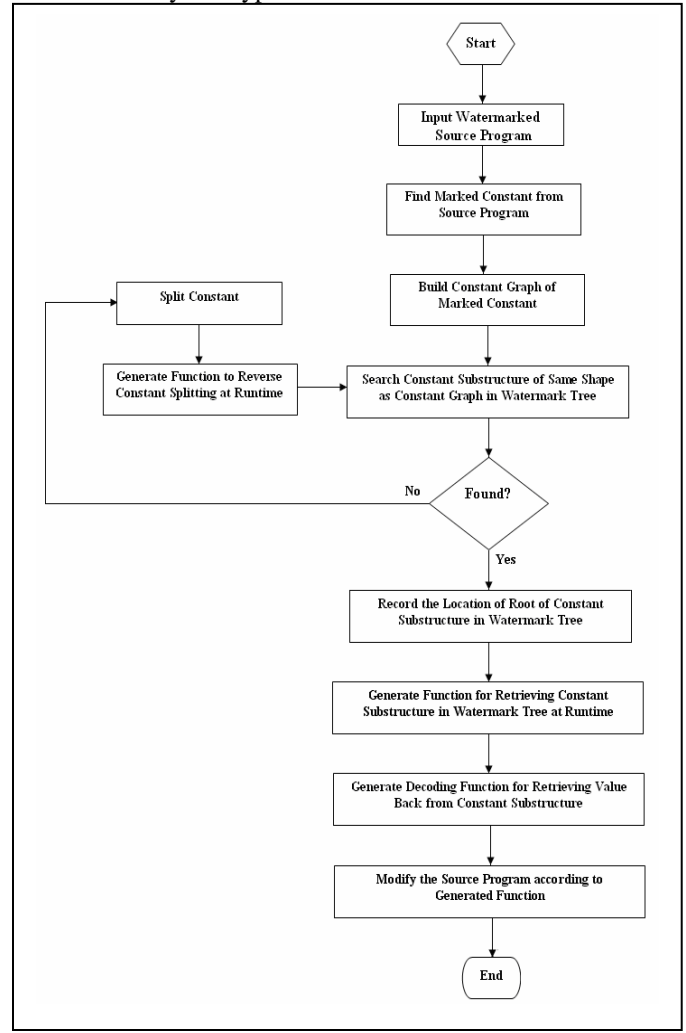

Figure 1.   Constant Encoding Technique.

The scope of constant encoding technique is limited to the single tree structure which is of planted plane cubic tree shape.

#### B. Framework/Model of Constant Splitting Technique

Block diagram mentioned in this section describes the procedure of constant splitting which is essential for handling the encoding of those constants which cannot be encoded due to mismatch of shape.

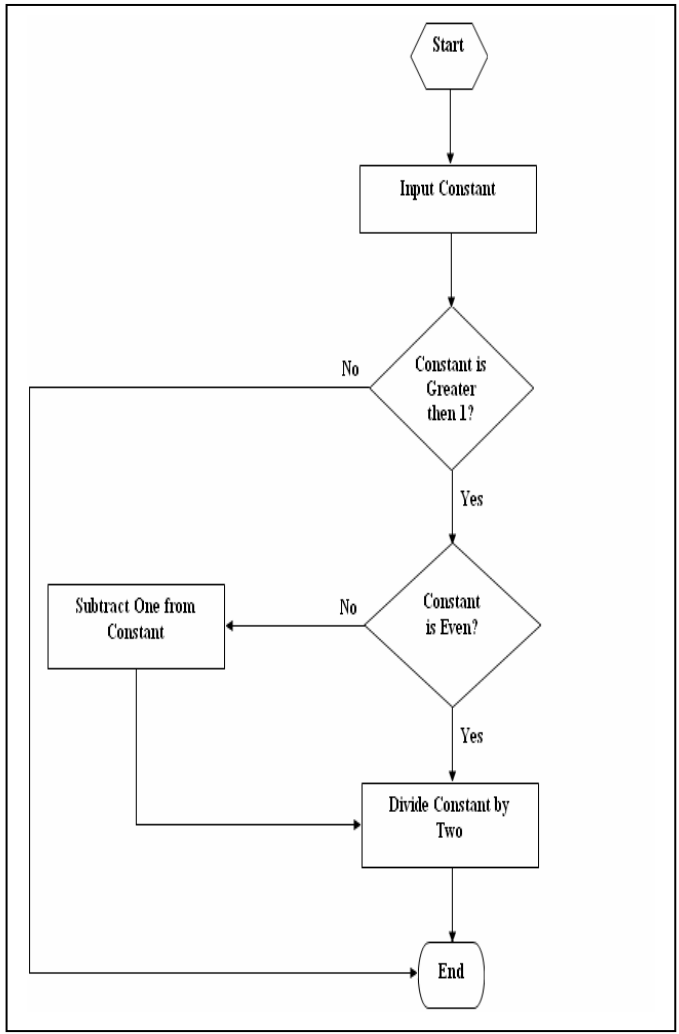

Figure 2. Constant Splitting Technique.

Constant splitting method is recursive in nature and applies until the constant substructure is not found in the watermark tree.

## IV. TECHNIQUE

### A. Algorithm of Constant Encoding Technique

- *Step 1:*

Input watermarked source program

- *Step 2:*

Select constants for encoding from the source program

- *Step 3:*

Build constant graph of the selected constants

- *Step 4:*

Search constant substructure of same shape as constant graph from watermark tree

- *Step 5:*

If constant substructure is not found in the watermark tree, then apply the constant splitting technique

- *Step 6:*

In the reverse situation, record the location of root of constant substructure in watermark tree

- *Step 7:*

Construct function to retrieve constant substructure in the watermark tree at runtime

- *Step 8:*

Generate decoding function to retrieve constant value back from the constant substructure

- *Step 9:*

Modify the watermarked source program according to generated decoding function

### B. Algorithm of Constant Encoding Technique

For the understanding of the algorithm, suppose that two variables namely as even and odd are taken having values '1' and '0' respectively

- *Step 1:*

Input constant value.

- *Step 2:*

Check that constant value is greater than '1'

- *Step 3:*

Check whether the constant is even or odd

- *Step 4:*

If constant is even, divide it by '2', and move to Step 6.

- *Step 5:*

If constant is odd, then first subtract '1' from it and then move to Step 7.

- *Step 6:*

Multiply the value of even by '2' and store the result again in even

- *Step 7:*

Add '1' in the value of odd and assign it again to odd, and then go to Step 4.

Our Technique is based on Thomborson's technique of constant encoding. The implementation and experimental results mentioned in next sections show that constant encoding can be done effectively with dynamic graph watermarking but with reasonable increase in code size, heap space usage and execution time. Our system is built in java language and its target language is also Java. The system takes the watermarked program as input and successfully outputs the constant encoded programs. For handling the encoding of large constant, constant splitting technique is used. The main purpose of constant splitting technique is that even if the whole constant could not encode due to dissimilarity of shapes, then some part of that of that will definitely encoded. Our implemented technique mainly does access the information of watermark tree provided in the watermark class of watermarked program. For referencing the constant substructure at the runtime in watermark tree, different mechanisms are employed.

## V. EPERIMENTAL RESULTS

We run the experiments under windows XP professionals with 512MB of RAM. The processor used in the experiments is Pentium 4. We use the J2SDK1.4.2_17 as backend java tool and Realj version 3.1 as front end tool. The system used for watermarking is SandMark v3.4.0, and decompilation process is done through Front End Plus v1.04.

We tested our system on different medium sized programs. Each program is watermark with the same three digit number. The number of encoded constants of watermarked program varied. Further description of the programs used for testing and analysis are given below in table 1

After constant encoding procedure, parameters which include code size, execution time, heap space usage and resilience are evaluated.

### A. Evaluation of Parameters

For the measurement of execution time, code size and heap space usage, each program is executed 'n' times. In evaluation of parameters, the comparison is made between the watermarked (WM) and tamper-proofed (TP) applications. Tables and Figures which are mentioned below specify the difference between the values of different parameters of watermarked and tamper-proofed applications along with the specified units used in measurement.

#### 1) Heap space usage:

*a) Tabular representation:* The table I given below distinguishes the usage of heap space after the constant encoding process of watermarked application.

| Program | Heap Space Usage | | |
|---|---|---|---|
| | *Watermark* | *Tamper-proof* | *Difference* |
| 1 | 359589.4 | 418581.2 | 58991.8 |
| 2 | 145648.0 | 200507.0 | 54859.0 |
| 3 | 278706.5 | 343980.9 | 68274.4 |
| - | - | - | - |
| - | - | - | - |
| - | - | - | - |
| - | - | - | - |
| - | - | - | - |
| n | 268397.2 | 254565.6 | 46168.4 |

a. Sample of a Table footnote. *(Table footnote)*

As the table indicates that no one application has the same incremented difference in heap space usage after constant encoding process.

*b) Graphical representation:* The graphical representation of Table 1 demonstrated in Figure 3 provides the conclusion that after constant encoding process, heap space usage of all watermarked application will definitely increase, but the ratio of increment.

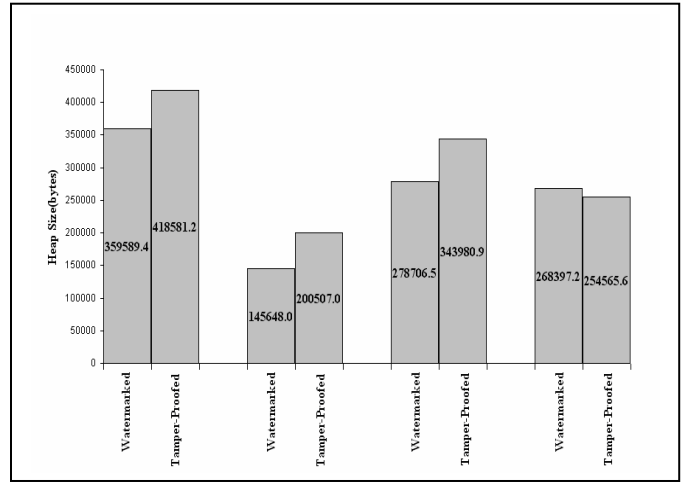

Figure 3. Difference in Heap Space Usage.

may vary due to dependency upon the number of constants encoded in each application. So less number of constants to be encoded will yield less increase in the size of heap space.

#### 2) Execution Time:

*a) Tabular representation:* Table II which is provided below illustrates the variation in execution time subsequent to the constant encoding process.

TABLE II. DIFFERENCE IN EXECUTION TIME

| Program | Execution Time | | |
|---|---|---|---|
| | *Watermark* | *Tamper-proof* | *Difference* |
| 1 | 339.405 | 451.900 | 12.405 |
| 2 | 60.550 | 69.175 | 8.625 |
| 3 | 323.485 | 329.755 | 6.27 |
| - | - | - | - |
| - | - | - | - |
| - | - | - | - |
| - | - | - | - |
| n | 119.840 | 127.115 | 7.275 |

a. Sample of a Table footnote. *(Table footnote)*

As the table reveals that after constant encoding process, execution time of each watermarked application is increased but proportion of change also differ here.

*b) Graphical representation:* Figure 4 depicts the conclusion that the dissimilarity in execution time of 'n' number of programs is due to the values of constants based on which constant graph is generated and the number of times the constant splitting technique is applied.

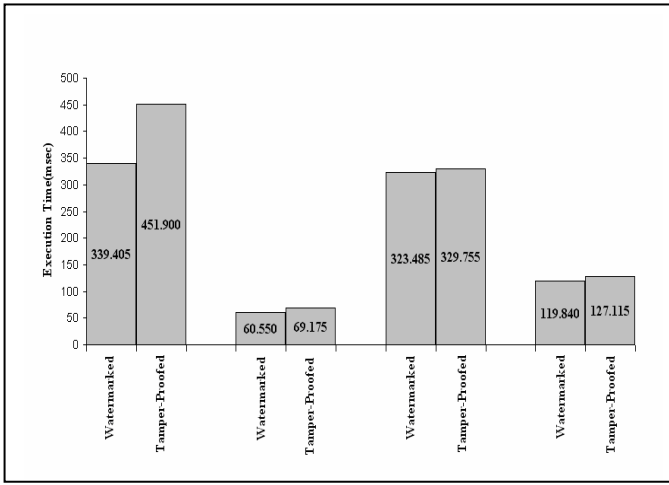

Figure 4. Difference in Execution Time.

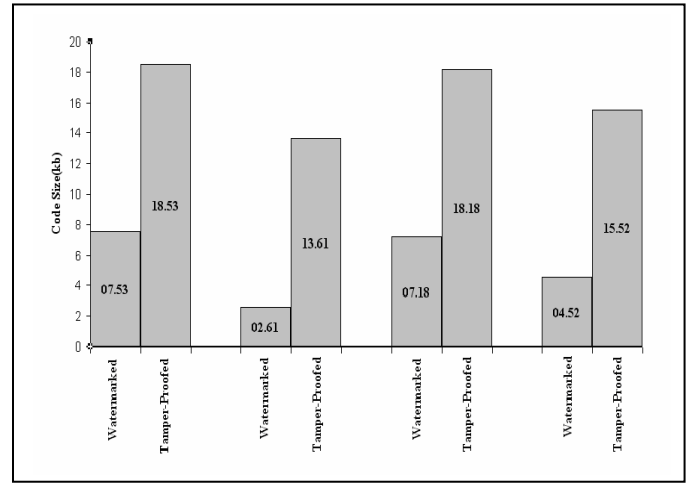

Figure 5. Difference in Code Size.

Thus it can be concluded that constant encoded applications will always have the more execution time then the watermarked application, but the percentage of increment may vary by reason of having total dependency upon the value of constant which will encode in the watermark tree.

### 3) Code Size:

*a) Tabular representation:* Table III gives an idea about the effect of constant encoding process on the code size of each watermarked program.

TABLE III.　　DIFFERENCE IN CODE SIZE

| Program | Code Size | | |
|---|---|---|---|
| | *Watermark* | *Tamper-proof* | *Difference* |
| 1 | 07.53 | 18.53 | 11.0 |
| 2 | 02.61 | 13.61 | 11.0 |
| 3 | 07.18 | 18.18 | 11.0 |
| - | - | - | - |
| - | - | - | - |
| - | - | - | - |
| - | - | - | - |
| - | - | - | - |
| n | 04.52 | 15.52 | 11.0 |

a. Sample of a Table footnote. *(Table footnote)*

As it is obvious from the table that after constant encoding process, all the applications has uniform increase in the code size.

*b) Graphical representation:* Figure 5 provides the graphical representation of disparity in code size which is mentioned in Table III. Analysis which is conducted on the basis of Table III and Figure 5 concludes that code size of each watermarked program is incremented by '11' kb and the cause of this augmentation is the amount of code inserted in watermarked program after the constant encoding process.

So number of constants and values of constants to be encoded will always have no effect on the code size of each application.

### 4) Resilience:
Resilience basically measures at what extent the watermarked and tamper-proofed application is unfeasible and invulnerable against transformation attacks such as semantic preserving transformations which include code obfuscation and code optimization. For evaluating the resilience level of watermarked and tamper-proofed application, techniques applied are class splitter, reorder instruction, duplicating the register, field assignment and variable reassigner

*a) Reorder Instruction Program Transformation Attack:* Reorder instruction attack tries to distort the program by reordering the instruction within each basic block of a method. Table IV indicates the consequences of reorder instruction attack and reveals that constant encoding process does not affect the resilience of watermarked application.

TABLE IV.　　EFFECT OF REORDER INSTRUCTION ATTACK

| Program | Reorder Instruction Attack | |
|---|---|---|
| | *Watermark* | *Tamper-proof* |
| 1 | Not affected | Not affected |
| 2 | Not affected | Not affected |
| 3 | Not affected | Not affected |
| - | - | - |
| - | - | - |
| - | - | - |
| - | - | - |
| - | - | - |
| n | Not affected | Not affected |

a. Sample of a Table footnote. *(Table footnote)*

*b)* *Class Splitter Program Transformatioon Attack:* This attack utilizes the technique of splitting class in half by placing some methods and fields to the super class. Table V given below illustrates that constant encoded applications are susceptible to class splitter attack in contradictory to watermarked applications.

TABLE V. EFFECT OF CLASS SPLITTER ATTACK

| Program | Class Splitter Attack | |
|---------|-----------|-------------|
| | *Watermark* | *Tamper-proof* |
| 1 | Not affected | Affected |
| 2 | Not affected | Affected |
| 3 | Not affected | Affected |
| - | - | - |
| - | - | - |
| - | - | - |
| - | - | - |
| - | - | - |
| n | Not affected | Affected |

a. Sample of a Table footnote. *(Table footnote)*

*c)* *Duplicate Register Program Transformation Attack:* Procedure of duplicate register involves in taking a local variable in a method and then split reference to it with a new variable. Table VI provides the detail about the successful behavior of the watermarked programs after constant encoding process after the attack.

TABLE VI. EFFECT OF DUPLICATE REGISTER ATTACK

| Program | Duplicate Register Attack | |
|---------|-----------|-------------|
| | *Watermark* | *Tamper-proof* |
| 1 | Not affected | Not affected |
| 2 | Not affected | Not affected |
| 3 | Not affected | Not affected |
| - | - | - |
| - | - | - |
| - | - | - |
| - | - | - |
| - | - | - |
| n | Not affected | Not affected |

a. Sample of a Table footnote. *(Table footnote)*

*d)* *Field Assignment Program Transformation Attack:* This attack performs obfuscation by inserting bogus field into class and then making assignment to that field in different locations of the program Table VII depicts the effect of field assignment attack and shows that this attack has no influence on the performance of the tamper-proofed applications.

TABLE VII. EFFECT OF FIELD ASSIGNMENT ATTACK

| Program | Field Assignment Attack | |
|---------|-----------|-------------|
| | *Watermark* | *Tamper-proof* |
| 1 | Not affected | Not affected |
| 2 | Not affected | Not affected |
| 3 | Not affected | Not affected |
| - | - | - |
| - | - | - |
| - | - | - |
| - | - | - |
| - | - | - |
| n | Not affected | Not affected |

a. Sample of a Table footnote. *(Table footnote)*

*e)* *Varaiable Reassigner Program Transformation Attack:* Variable reassigner functions in program by reallocating the local variable in order to minimize the use of number of local variable slots. Table VIII demonstrates that variable reassigner attack makes all the watermarked and tamper-proofed applications no more to executable and it also indicates no improvement in resilience level of watermarked application even after the constant encoding process against this attack.

TABLE VIII. EFFECT OF VARIABLE REASSIGNER ATTACK

| Program | Variable Reassigner Attack | |
|---------|-----------|-------------|
| | *Watermark* | *Tamper-proof* |
| 1 | Affected | Affected |
| 2 | Affected | Affected |
| 3 | Affected | Affected |
| - | - | - |
| - | - | - |
| - | - | - |
| - | - | - |
| - | - | - |
| n | Affected | Affected |

a. Sample of a Table footnote. *(Table footnote)*

The analysis which is conducted helps in concluding that the only thing that can affect the resilience of tamper-proofed application is large code size.

VI. CONCLUSION AND FUTURE WORK

The implementation of constant encoding tamper-proofing process provided in this paper is promising step for further research. The analysis which has been performed for evaluating the effectiveness of dynamic graph software watermarks concludes with considerable effect of tamper-proofing process on dynamic graph software watermarks. It is summarized from the evaluation of parameters that after constant encoding process code size of the application is always increased, but the incremental change in heap space usage and execution time may vary due to dependence upon various other factors. After constant encoding, the level of resilience of watermarked application can also be degraded due to large code size. To make the tamper-proofed code resilient against all types of attacks, one needs to improve the constant encoding process by inserting opaque predicated or through some other obfuscation techniques.

In this paper, constant splitting technique is adopted in combination with Clark Thomborson's technique to handle all the constants which are greater or smaller than the watermark value. Another better approach which can improve this integration of software protection techniques to great extent is that, instead of breaking up the constant value, splits the constant graph into subgraphs in situation where constant substructure is not found in watermark tree, though it would results in difficulty of tracking all the references related to the roots of subgraphs, but this method can provide high level of protection against attacks.

ACKNOWLEDGMENT